# HIGH STRENGTH OVER DENSITY RATIO INVISIBLE CABLES

Nicola M. Pugno
Department of Structural Engineering, Politecnico di Torino,
Corso Duca degli Abruzzi 24, 10129 Torino, Italy
nicola.pugno@polito.it

Spiders suggest us that producing high strength over density ratio invisible cables could be of great importance. In this letter we show that such invisible cables could in principle be built, thanks to carbon nanotube bundles. Theoretical strength of ~10MPa, Young's modulus of ~0.1GPa and density of $\sim 0.1 \text{Kg}/\text{m}^3$ are estimated.

Since their discovery [1,2], carbon nanotubes have stimulated intense study. In particular, unique and extraordinary mechanical properties were predicted [3-8], such as an extremely high Young's modulus (~1TPa), strength (~1GPa) and consequently failure strain (~0.1), similar to those of graphite in-plane [9]. Such properties have experimentally been confirmed by direct measurements performed by Ruoff and collaborators [10], who were able to develop a nanotensile testing apparatus by using two opposite atomic force microscope tips. Furthermore, the low carbon density (~1300Kg/m³) suggests that carbon nanotubes have promising high strength and lightweight material applications, e.g., for innovative nano-electromechanical systems [11-13]. In this letter we show that their predicted properties are sufficient to realize also invisible macroscopic cables.

Consider a rectangular cable having width $W$, thickness $T$ and length $L$; the cross-section be composed by $n \times m$ multiwalled carbon nanotubes with inner and outer diameter $d_-$ and $d_+$ respectively and length $L$. Let us assume that they are arranged in a square lattice with periodic spacing $p=W/n=T/m$. Then, the strength $\sigma_C$ of the cable, defined as the failure tensile force divided by the nominal area $W \times T$, is predicted as:

$$\sigma_C = \frac{\pi}{4} \frac{d_+^2 - d_-^2}{p^2} \sigma_{NT}, \quad \sigma \to E, \rho \qquad (1)$$

where $\sigma_{NT}$ denotes the strength of the single carbon nanotube. To derive eq. (1) we have assumed a full transfer load between the nanotube shells, that seems to be plausible if intertube bridgings are present [14]. The same relation is derived for the cable Young's modulus $E_C$ considering in eq. (1) the substitution $\sigma \to E$ and $E_{NT}$ as the Young's modulus of the single carbon nanotube. Similarly, the cable density $\rho_C$, defined as the cable weight divided by the nominal volume $W \times T \times L$, is predicted according to eq. (1) with the substitution $\sigma \to \rho$, where $\rho_{NT}$ would denote the carbon (nanotube) density. Thus, the same (failure) strain $\varepsilon_C = \sigma_C / E_C = \sigma_{NT} / E_{NT}$ and strength over density ratio $R = \sigma_C / \rho_C = \sigma_{NT} / \rho_{NT}$ is expected for the cable and for the single nanotube. Eq. (1) can be considered a general law to connect the nanoscale properties of the single nanotube with the macroscopic properties of the cable.

On the other hand, indicating with $\lambda$ the light wavelength, the condition for a nanotube to be invisible is:

$$d_+ \ll \lambda \tag{2a}$$

whereas to have a globally invisible cable, we require to have not interference between single nanotubes, i.e.:

$$p \gg \lambda \tag{2b}$$

We do not consider here the less strict limitations imposed by the sensitivity of the human eye, that can distinguish two different objects only if their angular distance is larger than ~1'. In other words, we want the cable to be intrinsically invisible.

Assuming $d_+/\lambda \approx 1/10$, $p/\lambda \approx 10$, to satisfy eqs. (2), the previously reported nanotube theoretical strength, Young's modulus and density, would yield, according to eq. (1), the following wavelength-independent invisible cable properties:

$$\sigma_C^{(theo)} \approx 10\text{MPa}, \ E_C \approx 0.1\text{GPa}, \ \rho_C \approx 0.1\,\text{Kg}/\text{m}^3 \tag{3}$$

Defects are expected to considerably reduce the strength of the nanotube bundle. For example, assuming large (with respect to the atomic spacing), holes a strength reduction by a factor of 3.36 is expected according to quantized fracture mechanics [15]. Atomistic simulations based on molecular mechanics or quantum mechanics (density functional theory) confirm such a prediction [16]. More critical defects could also be present in the cable, such as nanocracks [15]. However the lowest strength for carbon nanotubes measured in [13] was $\sigma_{NT} \approx 11\text{GPa}$, which would correspond to $\sigma_C \approx 1\text{MPa}$. Furthermore, considering $d_+/\lambda \approx 1/\sqrt{10}$ instead of $d_+/\lambda \approx 1/10$ would yield $\sigma_C^{(theo)} \approx 100\text{MPa}$. Thus, we conclude that invisible extremely lightweight cables with strength in the megapascal range could be realized, thanks to carbon nanotube technology.

Meter-long multiwalled carbon nanotube cables can already be realized [17]. For such a nanostructured macroscopic cable a strength over density ratio of $R = \sigma_C/\rho_C \approx 120-144\,\text{KPa}/(\text{Kg}/\text{m}^3)$ was measured, dividing the breaking tensile force by the mass per unit length of the cable (the cross-section geometry was not of clear identification). Thus, we estimate (eq. (1)) for the single nanotube contained in such a cable $\sigma_{NT} \approx 170\text{MPa}$ ($\rho_{NT} \approx 1300\,\text{Kg}/\text{m}^3$), much lower than its theoretical [16] or measured [17] nanoscale strength. This result was expected as a consequence of the larger probability to find critical defects in larger volumes [18]. Thus, defects could strongly limit the range of applicability of long bundles based on nanotubes. However, the cable strength is expected to increase with the technological advancement. The cable density was estimated to be $\rho_C \approx 1.5\,\text{Kg}/\text{m}^3$ [17], thus resulting in a cable strength of $\sigma_C \approx 200\text{KPa}$ (eq. (1)). Note that a densified cable with a larger value of $R = \sigma_C/\rho_C \approx 465\,\text{KPa}/(\text{Kg}/\text{m}^3)$ was also realized [17], suggesting the possibility of a considerable advancement for this technology in the near future. For such cables a degree of transparency was observed, confirming that our proposal is realistic. Inverting eq. (1) we deduce for them $p \approx 260\text{nm}$, in good agreement with the Scanning Electron Microscope (SEM) image analysis [17]. The nanotube characteristic diameter was $d_+ \approx 10\text{nm}$. Considering the visible spectrum, $\lambda \approx 400-600\text{nm}$, the condition (2a) was thus satisfied, even if the formation of fibrils of $d_+ \approx 50\text{nm}$ was observed in SEM [17], in contrast to the condition (2b). Thus, only a partial degree of transparency was to be expected.

Moreover, multiwalled carbon nanotubes with $d_+ \approx 50\text{nm}$ ($d_- \approx 0\text{nm}$) spaced by $p \approx 5\mu\text{m}$ are expected to realize an invisible cable with the properties given in eq. (3). For example this would correspond to a 10cm×1mm invisible and flexible cable with a weight per unit length of only $10\mu\text{g/m}$, capable of supporting the weight of a man (1000N).

A later force at the middle of the cable will induce a maximum strain $\varepsilon \approx 2(W/L)^2$ before all the nanotubes are in contact. Since for a cable $W/L \ll 1$ (e.g., $10^{-2}$) a strain of the order of $\varepsilon \approx 10^{-4}$, i.e., small if compared with that at failure $\varepsilon_{NT}^{(theo)} \approx \sigma_{NT}^{(theo)}/E_{NT} \approx 0.1$, will activate the nanotube interaction. In such a situation the cable would "appear" near the point of application of the lateral force, i.e., where the conditions of eqs. (2) is not locally verified, to survive and activate the lateral interaction between the nanotubes.

Summarizing, in this letter we have shown that high strength over density ratio invisible cables could be produced in the near future, thanks to carbon nanotube technology. However, defects could pose limitations to the realization of such cables. Even if further detailed studies are needed to improve the rough design proposed in this letter, the degree of transparency observed in meter-long carbon nanotube cables [17] seems to confirm the validity of our proposal.

Acknowledgements

The author would like to thank Profs. G. A. Pugno, A. Carpinteri and P. P. Delsanto for discussion.